\documentclass[amsmath,amssymb,onecolumn, a4paper,14pt]{revtex4}
\usepackage{amsmath}
\usepackage{amssymb}
\usepackage{amsmath,amsthm,amssymb,amscd}
\usepackage{latexsym}
\usepackage{indentfirst}
\usepackage{subfigure}
\usepackage{graphicx}
\usepackage{extarrows}
\usepackage{bm}
\usepackage{multirow}
\usepackage[colorlinks,linkcolor=blue,hyperindex,dvipdfm]{hyperref}

\date{\today}

\begin{document}

\title{
  \bf Relationship between promoter sequence and its strength in gene expression
}

\author{Jingwei Li and Yunxin Zhang} \email[Email: ]{xyz@fudan.edu.cn}
\affiliation{Laboratory of Mathematics for Nonlinear Science, Shanghai Key Laboratory for Contemporary Applied Mathematics, Centre for Computational Systems Biology, School of Mathematical Sciences, Fudan University, Shanghai 200433, China. }

\begin{abstract}
Promoter strength, or activity, is important in genetic engineering and synthetic biology. Evidences show that a constitutive promoter with certain strength for one given RNA can often be reused for other RNAs. Therefore, the strength of one promoter is mainly determined by its nucleotide sequence. One of the main difficulties in genetic engineering and synthetic biology is how to control the expression of certain protein in one given level. One usually used way to achieve this goal is to choose one promoter with suitable strength, which can be employed to regulate the rate of transcription and then leads to needed level of protein expression. For this purpose, so far, many promoter libraries have been established experimentally. However, theoretical methods to predict the strength of one promoter from its nucleotide sequence are desirable. Since such methods are not only valuable in the design of promoter with specified strength, but also meaningful to understand the mechanism of promoter in gene transcription. In this study, through various tests one theoretical model is presented to describe the relationship between promoter strength and its nucleotide sequence. Our analysis shows that, promoter strength is greatly influenced by nucleotide groups with three adjacent nucleotides in its sequence. Meanwhile, nucleotides in different regions of promoter sequence have different effects on promoter strength. Based on experimental data for {\it E. coli} promoters, our calculations indicate, nucleotides in -10 region, -35 region, and the discriminator region of promoter sequence are more important than those in spacing region for determining promoter strength. With model parameter values obtained by fitting to experimental data, four promoter libraries are theoretically built for the corresponding experimental environments under which data for promoter strength in gene expression has been measured previously.
\end{abstract}


\maketitle

\section{Introduction}
In cells, a small variety of expression of some protein may influence cell metabolism seriously. In synthetic biology, many models have been presented to describe the metabolic network \cite{Kitano2002,Yeang2006}. According to these models, it often needs to express a certain kind of protein (especially enzyme) in a specific intensity. One of the widely used ways to do this is to adjust the nucleotide sequence of the corresponding promoter \cite{Jensen1998,Mijakovic2005,Sanchez2011,Mutalik2013Reuse}.

\begin{figure}[!ht]
  \centering
  \includegraphics[width=15cm]{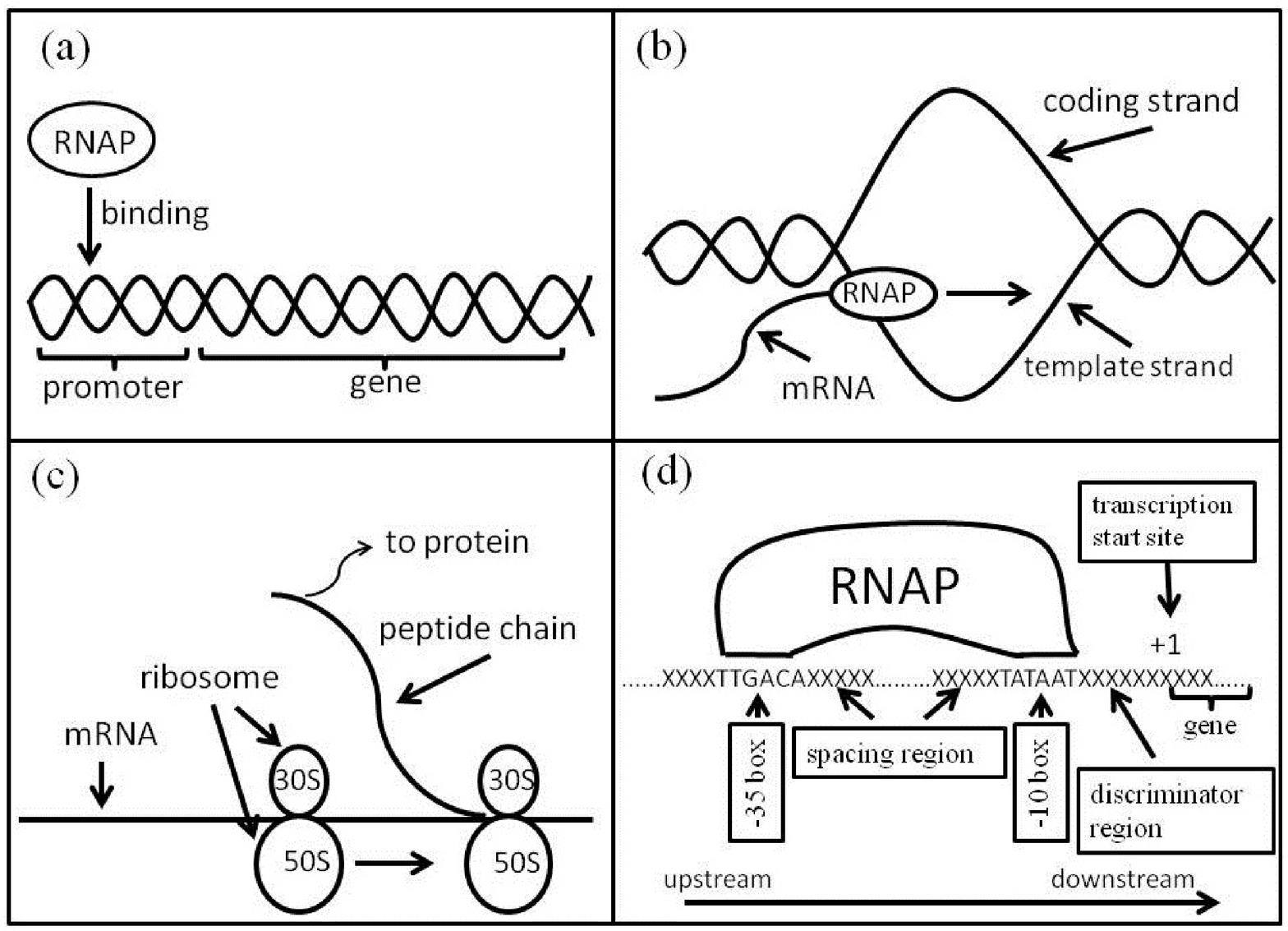}\\
  \caption{Schematic depiction of gene expression, including transcription process and translation process. The transcription is initiated by the binding of RNA polymerase to promoter {\bf (a)}. With the motion of RNAP along unpaired DNA, message RNA (mRNA) is then produced with the genetic information coded in DNA {\bf (b)}. During the translation process, the previously obtained mRNA is then used to assemble amino acids into protein {\bf (c)}. The rate of protein production can be regulated by the sequence of promoter. In {\it E. coli}, promoter sequence contains two short sequence elements approximately -10 and -35 nucleotides upstream from the transcription start site. The sequence at -10 (called -10 region or -10 box) has the consensus sequence TATAAT. The sequence at -35 (called -35 region or -35 box) has the consensus sequence TTGACA  {\bf (d)}.
  }\label{FigSchematic}
\end{figure}

A promoter is a region of DNA that initiates transcription of a particular gene \cite{Matsuyama2004Search,Mulligan1984}, see Fig. \ref{FigSchematic}. In gene expression, the genetic information coded in nucleotide sequence of DNA should be firstly transcribed into message RNA (mRNA), which is performed by enzyme RNA polymerase (RNAP) \cite{Gross1996,Dehaseth1998}. Usually, the transcription process begins with the binding of RNAP to one specific upstream region of the target gene, which is called promoter \cite{Mishra1993,Campbell2002}. Experiments show that, with different promoters, the protein production rates, or the strengths of gene expression, will be different \cite{Jensen1998,Lu2012,Mutalik2013Reuse}. Therefore, the rate of gene transcription to mRNA is regulated by the nucleotide sequence of promoter (for simplicity, in this study, promoters are assumed to be constitutive, i.e. transcription rate of the corresponding downstream gene is not influenced by transcription factors, for related discussions about translation factors one can see \cite{Bintu2005Regulatorymodel,Kinney2010Regulatory}). Due to the requirement of genetic engineering and synthetic biology, the production rates of certain proteins, especially some enzymes, should be regulated detailedly. One of the ways to attain this aim is to choose specific promoter sequence to get needed rate of transcription (another efficient way is to choose specific ribosome binding site sequence, i.e. RBS sequence, to regulate the rate of translation \cite{Salis2009,Mutalik2013Reuse}).

In order to achieve this goal, many promoter libraries corresponding to large-scale strength of gene expression have been built experimentally \cite{Jensen1998,Mey2007LibraryNearby,Alper2005LibraryReuse,Ruhdal1998LibrarySpacer,Rud2006Library,Solem2002Library,Rhodius2012,Lu2012,Wu2013}. Since it has been experimentally verified that the activity of a promoter can be reused among different kinds of proteins \cite{Alper2005LibraryReuse,Mutalik2013Reuse,Cox2007Reuse,Barnard2007Reuse}, these promoter libraries are valuable to the regulations of metabolic networks involving many different kinds of proteins. The establish of promoter libraries is of great important to the development of synthetic biology. However, reliable mathematical models to describe the relationship between promoter strength and its nucleotide sequence are much desirable. Since such models can not only reduce the experimental expense in building further promoter libraries, but also can help us to understand the mechanism of promoter during gene transcription. More importantly, such models will make it easier to get needed promoter with specific expression strength, and then will be valuable in genetic engineering and synthetic biology.

In the last twenty years, many studies have been done to build quantitative relationship between promoter strength in gene expression and its nucleotide sequence. It has been discovered early that, in {\it E. coli} promoters, the -10 region hexamer and -35 region hexamer are strongly conserved, and they are much important for determining the expression strength of promoter \cite{Hawley1983Nature,Dehaseth1998,Djordjevic2011}. But recent experimental data shows that promoter strength also depends on nucleotide types in the spacing region of promoter sequence \cite{Wu2013}, and may also depend on the discriminator region. Where spacing region is the promoter sequence region between the most conserved -10 region and -35 region, and discriminator region is the sequence region between the -10 region and transcription start site, see Fig. \ref{FigSchematic}(d).
Therefore, it becomes much difficult to build reasonable theoretical models to describe the relationship between promoter sequence and its strength, since there may be too many factors or variables which may affect the promoter strength.

In \cite{Rhodius2010OtherPredict}, one modular position weight matrix model is presented to evaluate the contribution of promoter sequence to its strength. Where promoter score, which correlates with protein-DNA binding energy and consequently correlates with promoter strength, is obtained as one linear combination of scores for each active promoter sequence modular and an additional penalty term for nonoptimal modular, with the sequence modular scores obtained by basic principles of statistical physics. Similar idea has also been used in \cite{Mulligan1984,Berg1987Regulatory} to try to understand the promoter strength from its nucleotide sequence. Meanwhile, in \cite{Kinney2010Regulatory}, based on a large number of experimental data for strengths and sequences of {\it E. coli lac} promoters, an adapted energy matrix for RNAP binding to promoter is statistically determined. Based on this energy matrix, one thermodynamic model is designed in \cite{Brewster2012Plos} to predict promoter strength from its nucleotide sequence. On the other hand, by using support vector regression method and the distribution of specific nucleotides at each position of promoter sequence, which is obtained in \cite{Hawley1983Nature} and based on 168 {\it E. coli} promoter sequences, one strength prediction skill of {\it E. coli} promoter from microarray data is provided in \cite{Kiryu2005Support}. They found that several non-consensus nucleotides in the -10 region and -35 region of promoter sequences act positively on the promoter strength, while certain consensus nucleotides have only a minor effect on the strength.

Although many related studies have shown that the consensus sequences (-35 and -10 regions) are most essential to determine the strength of promoter \cite{Cheetham1999RNApromoter,Buc1985RNApromoter,Ross1993RNApromoter,Ruhdal1998LibrarySpacer}, and actually for the sake of simplicity most of the existing models are based on such assumption, recent experimental data for promoter strength obtained by Wang's study group in \cite{Wu2013} indicates that nucleotide types in spacing region of promoter sequence are also not neglectable. With different spacing sequences but keeping -10 region and -35 region unchanged, the expression strength of promoter may vary between 31 and 105 (in unit relative intensity of red fluorescent per OD$_{600}$, i.e. RIRF/OD$_{600}$). For simplicity, this study assumes that gene transcription will initiate as soon as one RNAP binds to its upstream promoter, and the nucleotide sequence of gene is not too long. Then for low concentration of RNAP, which assures that the elongation of mRNA (i.e. the motion of RNAP along DNA) will not be jammed \cite{Zhang2012,Fange2014}, the transcription rate will mainly be determined by the binding rate of RNAP to promoter. Biophysically, RNAP binding rate to promoter can be roughly written as $k=k_0e^{-\Delta G/k_BT}$. Where $k_0$ is one rate constant which depends on RNAP concentration and other experimental environments, $k_B$ is Boltzmann constant, $T$ is the absolute temperature, and $\Delta G$ is the free energy barrier of RNAP binding to promoter which is mainly determined by the promoter sequence. Generally, the energy barrier $\Delta G$ may depend on the secondary structure of promoter, and may also depend on the concentrations of transcription factors for nonconstitutive promoters. But this study assumes that $\Delta G$ can be completely determined by the nucleotide sequence of promoter, and the main focus in the following is to find one reasonable method to get energy barrier $\Delta G$ from the nucleotide sequence of promoter.

One immediate idea to get energy barrier $\Delta G$ is to assume that $\Delta G$ can be approximated by one linear combination of energy barriers $\Delta G_i$ contributed by each nucleotide in the promoter sequence. Here $i$ is the index of  position of nucleotide in promoter sequence. This idea is similar as the ones previously used in \cite{Mulligan1984,Berg1987Regulatory,Djordjevic2008,Rhodius2010OtherPredict}. The main difference between the idea here and the previous ones is as follows. In the previous methods, such as the one used in the modular position weight matrix model in \cite{Rhodius2010OtherPredict}, $\Delta G$ is assumed to be determined only by some so called active modules, including -35 region, -10 region, discriminator region, transcription start region, and the contribution from other suboptimal regions is only included as one penalty term which depends only on their lengths of nucleotide sequences. As have been mentioned previously, recent experimental data presented in \cite{Wu2013} indicates promoter strength also changes with the nucleotide types in spacing region. So we need to calculate the energy barrier contribution $\Delta G_i$ from nucleotide in any position $i$ of promoter sequence.
There are several possible ways to get $\Delta G_i$, which base more or less on basic principles of statistical physics \cite{Berg1987,Stormo1990,Rhodius2010OtherPredict}. For example, the energy barrier contribution of one nucleotide at position $i$ with type $b$ ($b=A, T, G$ or $C$) can be obtained as $\Delta G_{b,i}=\ln((n_{b,i}+0.005N)/(N+0.02N)/p_b)$ \cite{Rhodius2010OtherPredict}. Where $n_{b,i}$ is the number of nucleotide $b$ at position $i$ in the aligned sequences, $N$ is the total number of promoter, and $p_b$ is the theoretical probability of finding each type of nucleotide (usually $p_b=0.25$ is used), and 0.005 and 0.02 are just two other model parameter values. Such method may be theoretically sound, but our calculations indicate it is not reasonably good to describe the experimental data of promoters in \cite{Wu2013}.

If the types of nucleotide at different positions of promoter sequence are independent to each other, then due to principles of statistical physics, the probability that there is one base $b$ at position $i$ is related to the free energy $\Delta G_{b,i}$ by $p_{b,i}=\exp(-\Delta G_{b,i}/k_BT)$. By replacing probability $p_{b,i}$ with frequency $n_{b,i}$, we then get $\Delta G_{b,i}\approx -k_BT\ln n_{b,i}$. Where the frequency $n_{b,i}$ is obtained from the 168 promoters of {\it E. coli} compiled in \cite{Hawley1983Nature}. The total energy barrier $\Delta G$ for one given promoter sequence can then be obtained by $\Delta G=\sum \Delta G_{b,i}$, with $b$ the nucleotide at position $i$. However, this model can not fit to the recent experimental data presented in \cite{Wu2013} well, see Fig. S1(a) in Supplemental Material \cite{supplemental}.

In recent paper \cite{Kinney2010Regulatory}, based on experimental data of {\it E. coli lac} promoters, one $4\times75$ parameter matrix $M$ is built, which is used to describe the interaction of RNAP with promoter region from position -1 to position -75. Where the matrix element $M_{b,i}$ represents the contribution to this interaction from having a base $b$ at position $-i$ in the promoter sequence. Using this parameter matrix $M$, the total energy barrier for RNAP binding to one promoter can be known, and so the corresponding expression strength can be obtained. This parameter matrix has been employed by Brewster {\it et al.} in \cite{Brewster2012Plos} to build one theoretical model to predict the strength of promoter, and its accuracy has been validated in the design of promoter with specific strength. However, our test shows that such method is not satisfying when it is employed to describe the recent experimental data presented in \cite{Wu2013}, see Fig. S1(b) in Supplemental Material \cite{supplemental}.

One common characteristic of the three methods discussed above is that, the energy barrier $\Delta G$ of RNAP binding to promoter is assumed to be one linear combination of energy barriers $\Delta G_{b,i}$ contributed by each nucleotide in promoter sequence. The failures of these methods in describing the recent experimental data presented in \cite{Wu2013} implie that this assumption of independence and additivity of energy barriers may not be generally true, though it may be approximately reasonable in some special cases and has been validated previously for transcription factor binding sites of promoter \cite{Benos2002}. Therefore, this assumption needs to be modified to be more reasonable. One immediate way to do this is to assume that the energy barrier $\Delta G$ of RNAP binding to promoter is one linear combination of the ones  contributed by all nearest-neighbor (NN) nucleotide groups in promoter sequence. For convenience, this study assumes that the energy barrier contributed by one NN nucleotide group is equal to the energy obtained by {\bf NN model} in the study field of nucleic acids \cite{JOHN1998}. Which has actually been used in some software packages, such as NUPACK \cite{Zadeh2011}, to calculate the folding free energy of nucleic acid sequence to determine its secondary structure. In {\bf NN model}, the total energy of one given nucleic acid sequence is obtained as the summation of energies contributed by each NN nucleotide groups, and one additional term according to the initial nucleotide, see the corresponding values listed in Fig. S1(d) \cite{supplemental}.  For example, the energy of sequence CGTTGA at temperature 37$^o$C is obtained as $\Delta G=\Delta G({\rm CG})+\Delta G({\rm GT})+\Delta G({\rm TT})+\Delta G({\rm TG})+\Delta G({\rm GA})+\Delta G({\rm init.})=-2.17-1.44-1.00-1.45-1.30+(0.98+1.03)$ (excerpted from Ref. \cite{JOHN1998}). With one additional parameter to indicate the average level of experimental environments, the fitting results of this model are plotted in Fig. S1(c) \cite{supplemental}. Unfortunately, this model is also not satisfactory.

The failure of the above NN method to predict promoter strength, i.e. the failure in  calculating energy barrier $\Delta G$ of RNAP binding to promoter, may due to following two reasons. (1) In the NN method, no difference is included among energy contributions from nucleotide groups in -10 region, -35 region, discriminator region, and spacing region. As mentioned above, previous studies have shown that the -10/35 region may be more important for determining promoter strength. So the nucleotide (group) in -10/-35 region should contribute more to the RNAP binding energy $\Delta G$.  (2) The energy barrier $\Delta G$ may also depend on large nucleotide groups, at least on nucleotide groups with three adjacent nucleotides. In the following, we will test new models which do not have at least one of the above two weaknesses.

Firstly, we test the model which includes nucleotide position explicitly, where the energy barrier $\Delta G$ of RNAP binding to promoter is assumed to be linear combination of $\Delta G_{b,i}$ and $\Delta G_{b\bar{b},i}$. Here $\Delta G_{b,i}$ is the energy barrier contributed by one nucleotide $b$ at position $i$, and $\Delta G_{b\bar{b},i}$ is the energy barrier contributed by one nearest-neighbor nucleotide group $b\bar{b}$ with nucleotide $b$ at position $i$ and nucleotide $\bar{b}$ at position $i+1$. For convenience, in the following, this model is called POSITION2 model. One main difficulty in such model is, compared with the known experimental data, there are too many model parameters. For example, if the promoter sequence comprises of 35 nucleotides, then there will be $35\times4+34\times4^2=684$ model parameters. To avoid the overfitting problem, we used the partial least squares (PLS) regression to get parameter values. Where the principal component number, i.e. the number of independent model parameters, is chosen by 10-fold cross-validation. With the chosen principal component number, both the mean residual of all promoter strength between measured values and theoretical values, and the mean residual in 10-fold cross-validation are reasonably low, see Fig. S2(a,b) in Supplemental Material \cite{supplemental}. In this study, the experimental data includes a total of 422 promoters, with their nucleotide sequences and strengths presented in Refs. \cite{Lu2012,Wu2013,Mutalik2013Reuse} respectively. Due to the different experimental environments used in measuring promoter strengths, including temperature, RNAP type and concentration, and lots of other conditions in transcription and translation processes, in this model as well as other models used in the study, extra constants are added to total energy barrier $\Delta G$ to stand for these differences. For experimental data from different references, these extra constants will be different, and their values are also obtained by PLS regression. Meanwhile, to know if there are real differences between the energy barrier contributions from nucleotides in -10 region, -35 region, discriminator region, and spacing region, another extra constant is added to distinguish the length of promoter spacing region. The possible lengths of spacing region in the experimental data used in this study are 16, 17 and 18. With seven principal components, the fitting results of the POSITION2 model are presented in Fig. S2(c) \cite{supplemental}. Where the mean residual between the experimental data and theoretical values is around 61 (in arbitrary unit), see Fig. S2(a) in Supplemental Material \cite{supplemental}. From the model coefficient values plotted in Fig. S2(d,e,f) \cite{supplemental}, one can see that the contributions to total energy barrier $\Delta G$ from -10 region, -35 region, and discriminator region are larger than those from the spacing region for promoters with any length of spacing region. Here the model coefficients are obtained by an inverse transform from the values of the seven principal components.

One may argue that one of the reasons that the above NN model is not good enough to describe the relationship between promoter strength and sequence is that, the values of $\Delta G_{b\bar{b}}$ given by Ref. \cite{JOHN1998} for energy barrier contribution from nearest-neighbor nucleotide group $b\bar{b}$ may not be accurate enough (see also the values listed in Fig. S1(d) in Supplemental Material \cite{supplemental}), or they may not be generally right for any nucleic acid sequence. In order to exclude this reason, we have tested one generalized model, called GROUP2 model, in which the energy barrier $\Delta G$ of RNAP binding is assumed to be one linear combination of $\Delta G_{b}$ and $\Delta G_{b\bar{b}}$. Different with the above NN model, $\Delta G_{b}$ and $\Delta G_{b\bar{b}}$ here are obtained by PLS regression. However, from the plots in Fig. S2(a,b) \cite{supplemental}, one can see that the GROUP2 model is actually less accurate than the above POSITION2 model.

As mentioned previously, another possible reason for the failure of NN model to describe the relationship between promoter sequence and strength is that, the promoter strength may also depend on large nucleotide groups (with at least three adjacent nucleotides), but not only on independent nucleotides and nucleotide groups with two neighboring nucleotides. We may need to point out that, models with nucleotide groups of size only 3 implicitly include the cases of size 1 and 2. But, for the sake of comparison of contributions to the total energy barrier $\Delta G$ of RNAP binding to promoter, from nucleotide groups with different sizes, our model includes all possible  nucleotide groups with size 1, 2, and 3. For convenience, such model is called GROUP3 model. In GROUP3 model, there are also too many unknown model parameters which need to be fitted from experimental data, altogether $4+4^2+4^3=84$. Therefore, PLS regression is also used in the data fitting process to determine model parameter values, in which the principal component number (i.e. the number of independent model parameters) is determined by reducing both the mean residual between promoter strengths from experimental data and from theoretical model, and the mean residual of promoter strengths in 10-fold cross-validation, see Fig. S3(a,b) in Supplemental Material \cite{supplemental}. With eight principal components, the fitting results are presented in Fig. S3(c) \cite{supplemental}. From the model coefficient values plotted in Fig. S3(d) \cite{supplemental}, one can see that contributions to total energy barrier $\Delta G$ of RNAP binding from nucleotide groups with three adjacent nucleotides are relatively larger than those from nucleotides or nucleotide groups with two neighboring nucleotides.

\begin{figure}[!ht]
  \centering
  \includegraphics[width=15cm]{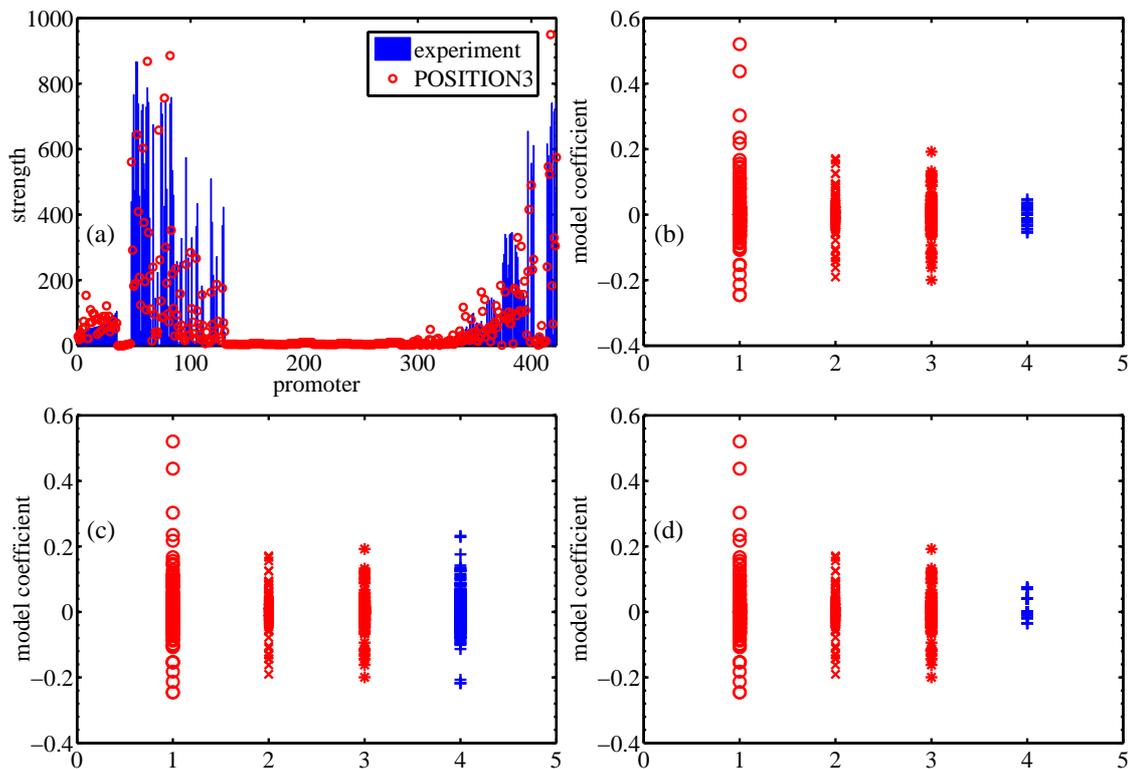}\\
  \caption{(Color online) {\bf (a)} Experimental data and theoretical predictions by POSITION3 model for promoter strength. There are altogether 422 promoters obtained in \cite{Lu2012,Wu2013,Mutalik2013Reuse} respectively, see also the description in Fig. S1 \cite{supplemental}. The parameter values (or model coefficients) of POSITION3 model are obtained by PLS regression with seven principal components, see Fig. S3 \cite{supplemental}. To show the difference of nucleotide contributions to promoter strength between nucleotides (or nucleotide groups) in spacing region of promoter sequence and nucleotides (or nucleotide groups) in other regions, the fitted parameter values of POSITION3 model are plotted in {\bf (b,c,d)}. Where the data points drawn at horizontal coordinates 1, 2, 3, 4 are corresponding to nucleotides in -35 region, -10 region, discriminator region, and spacing region respectively [see Fig. \ref{FigSchematic}(d)]. The lengths of promoter spacing region in {\bf (b,c,d)} are 16, 17, and 18 respectively. The plots in {\bf (b,c,d)} imply that, generally nucleotides in -10/-35 region and discriminator region are more important for determining the strength of promoter, see also Fig. S2(d,e,f) \cite{supplemental}.
  }\label{FigPosition3}
\end{figure}

Based on the above two test models, the POSITION2 model and the GROUP3 model, we  conclude that the energy barrier $\Delta G$ of RNAP binding to promoter depends on nucleotide groups with three adjacent nucleotides. Meanwhile, the contributions of nucleotide (groups) to $\Delta G$ from different regions of promoter sequence are different. Therefore, it seems more reasonable to use one combined model to describe the relationship between promoter strength and sequence, in which both nucleotide groups with size up to three and their positions in promoter sequence are explicitly considered. Undoubtedly, there will be too many unknown parameters in such combined models. For example, if each promoter has 35 nucleotides and there is no missing nucleotide in promoter sequence, the number of model parameters will be  $35\times4+34\times4^2+33\times4^3=2796$. Meanwhile, as mentioned previously, during data fitting extra parameters are also added due to the difference of length of promoter spacing region and the difference of experimental environment to measure promoter strength. Therefore, PLS regression should also be used to determine the values of model parameters. With reasonable low value of mean residual between theoretical model predictions and experimental data of promoter strengths, and mean residual of 10-fold cross-validation, the principal component number of this combined model is chosen to be seven, see Fig. S3(a,b) in Supplemental Material \cite{supplemental}. With these seven principal components, the theoretical results of this combined model (for convenience, this model is called POSITION3 model) are plotted in Fig. \ref{FigPosition3}(a). The mean residual between theoretical predictions and experimental measurements of promoter strengths is about 59 (in arbitrary unit), see Fig.S3(a). From the plots in Fig. \ref{FigPosition3}(b,c,d), one can easily see that the nucleotide groups in -10 region, -35 region, and discriminator region have more effects on promoter strength. Meanwhile, the effects of nucleotide groups with three adjacent nucleotides are also nonnegligible, see Fig. S4 in Supplemental Material \cite{supplemental}.

Using the POSITION3 model and genetic algorithm, four promoter libraries are calculated (see Fig. \ref{FigLibraries1}, and the Excel files in supplement materials for the corresponding promoter sequences), which correspond to the four experimental environments under which the data used in this study are obtained \cite{Lu2012,Wu2013,Mutalik2013Reuse}. These libraries will be helpful for the synthesis of promoter with specific expression strength. Using the same methods, promoter libraries corresponding to any other experimental environments can also be theoretically built, but with one extra step during which the model constant corresponding to the given experimental environment should be determined firstly by initial promoter samples. Combining our model here for promoter strength with the calculation method for ribosome binding site sequence \cite{Salis2009,Mutalik2013Reuse}, the gene expression, including transcription and translation, can be regulated to given strength detailedly.

In conclusion, in this study one theoretical model for describing the relationship between promoter strength and sequence is presented. Our study shows that the nucleotides in -10 region, -35 region, and the discriminator region have more effects on promoter strength than those in spacing region. Meanwhile, promoter strength depends more on nucleotide groups with three adjacent nucleotides than on single nucleotides and nucleotide groups with two neighboring nucleotides. Using our model, promoter libraries with wide range of expression strength are theoretically obtained.

\begin{acknowledgments}
This study was supported by the Natural Science Foundation of China (Grant No. 11271083), and the National Basic Research Program of China (National \lq\lq973" program, project No. 2011CBA00804).
\end{acknowledgments}


\newpage

\begin{figure}[!ht]
  \centering
  \includegraphics[width=15cm]{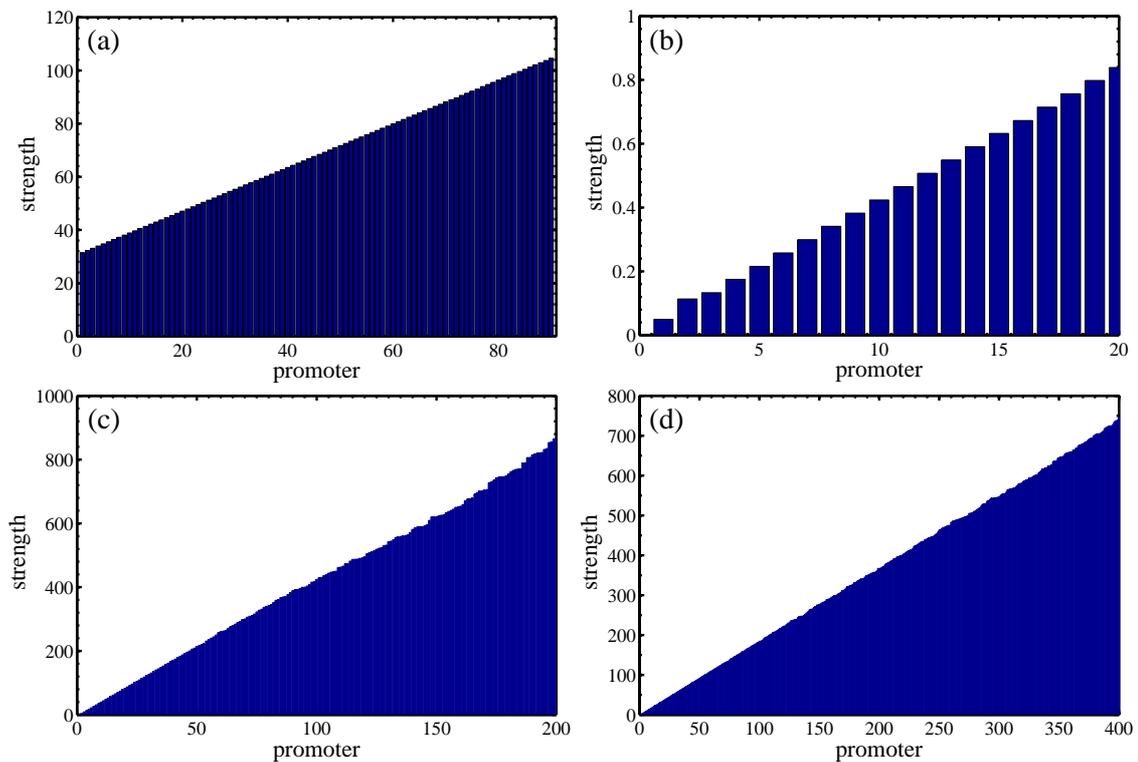}\\
  \caption{(Color online) Promoter libraries theoretically obtained by the POSITION3 model and genetic algorithm. Where the model parameter values are obtained by fitting to the experimental data measured in \cite{Lu2012,Wu2013,Mutalik2013Reuse}. {\bf (a)} 90 promoters with expression strength between 31-104 with the same measuring environment as in \cite{Wu2013}.  {\bf (b)} 20 promoters with one of the measuring environments in \cite{Lu2012} to build their promoter library. {\bf (c)} 200 promoters with the same measuring environment as in \cite{Mutalik2013Reuse} to build their modular promoter library (MPL), and {\bf (d)} 400 promoters with the same measuring environment as in \cite{Mutalik2013Reuse} to build their randomized promoter library (RPL).
  }\label{FigLibraries1}
\end{figure}

\end{document}